\documentstyle[twoside,epsf,amsmath,amssymb]{article}

\catcode`\@=11
\long\def\@makefntext#1{
\protect\noindent \hbox to 3.2pt {\hskip-.9pt  
$^{{\eightrm\@thefnmark}}$\hfil}#1\hfill}               

\def\thefootnote{\fnsymbol{footnote}}
\def\@makefnmark{\hbox to 0pt{$^{\@thefnmark}$\hss}}    
        
\def\ps@myheadings{\let\@mkboth\@gobbletwo
\def\@oddhead{\hbox{}
\rightmark\hfil\eightrm\thepage}   
\def\@oddfoot{}\def\@evenhead{\eightrm\thepage\hfil
\leftmark\hbox{}}\def\@evenfoot{}
\def\sectionmark##1{}\def\subsectionmark##1{}}

\oddsidemargin=\evensidemargin
\renewcommand{\thefootnote}{\fnsymbol{footnote}}

\newcounter{sectionc}\newcounter{subsectionc}\newcounter{subsubsectionc}
\renewcommand{\section}[1] {\vspace{12pt}\addtocounter{sectionc}{1} 
\setcounter{subsectionc}{0}\setcounter{subsubsectionc}{0}\noindent 
        {\tenbf\thesectionc. #1}\par\vspace{5pt}}
\renewcommand{\subsection}[1] {\vspace{12pt}\addtocounter{subsectionc}{1} 
        \setcounter{subsubsectionc}{0}\noindent 
        {\bf\thesectionc.\thesubsectionc. {\kern1pt \bfit #1}}\par\vspace{5pt}}
\renewcommand{\subsubsection}[1] {\vspace{12pt}\addtocounter{subsubsectionc}{1}
        \noindent{\tenrm\thesectionc.\thesubsectionc.\thesubsubsectionc.
        {\kern1pt \tenit #1}}\par\vspace{5pt}}
\newcommand{\nonumsection}[1] {\vspace{12pt}\noindent{\tenbf #1}
        \par\vspace{5pt}}

\newcounter{appendixc}
\newcounter{subappendixc}[appendixc]
\newcounter{subsubappendixc}[subappendixc]
\renewcommand{\thesubappendixc}{\Alph{appendixc}.\arabic{subappendixc}}
\renewcommand{\thesubsubappendixc}
        {\Alph{appendixc}.\arabic{subappendixc}.\arabic{subsubappendixc}}

\renewcommand{\appendix}[1] {\vspace{12pt}
        \refstepcounter{appendixc}
        \setcounter{figure}{0}
        \setcounter{table}{0}
        \setcounter{lemma}{0}
        \setcounter{theorem}{0}
        \setcounter{corollary}{0}
        \setcounter{definition}{0}
        \setcounter{equation}{0}
        \renewcommand{\thefigure}{\Alph{appendixc}.\arabic{figure}}
        \renewcommand{\thetable}{\Alph{appendixc}.\arabic{table}}
        \renewcommand{\theappendixc}{\Alph{appendixc}}
        \renewcommand{\thelemma}{\Alph{appendixc}.\arabic{lemma}}
        \renewcommand{\thetheorem}{\Alph{appendixc}.\arabic{theorem}}
        \renewcommand{\thedefinition}{\Alph{appendixc}.\arabic{definition}}
        \renewcommand{\thecorollary}{\Alph{appendixc}.\arabic{corollary}}
        \renewcommand{\theequation}{\Alph{appendixc}.\arabic{equation}}
        \noindent{\tenbf Appendix \theappendixc #1}\par\vspace{5pt}}
\newcommand{\subappendix}[1] {\vspace{12pt}
        \refstepcounter{subappendixc}
        \noindent{\bf Appendix \thesubappendixc. {\kern1pt \bfit #1}}
        \par\vspace{5pt}}
\newcommand{\subsubappendix}[1] {\vspace{12pt}
        \refstepcounter{subsubappendixc}
        \noindent{\rm Appendix \thesubsubappendixc. {\kern1pt \tenit #1}}
        \par\vspace{5pt}}

\newcommand{\textlineskip}{\baselineskip=13pt}
\newcommand{\smalllineskip}{\baselineskip=10pt}

\def\eightcirc{
\begin{picture}(0,0)
\put(4.4,1.8){\circle{6.5}}
\end{picture}}
\def\eightcopyright{\eightcirc\kern2.7pt\hbox{\eightrm c}} 

\newcommand{\copyrightheading}[1]
        {\vspace*{-2.5cm}\smalllineskip{\flushleft
        {\footnotesize International Journal of Modern Physics A, #1}\\
        {\footnotesize $\eightcopyright$\, World Scientific Publishing
         Company}\\
         }}

\def\abstracts#1#2#3{{
        \centering{\begin{minipage}{4.5in}\baselineskip=10pt\footnotesize
        \parindent=0pt #1\par 
        \parindent=15pt #2\par
        \parindent=15pt #3
        \end{minipage}}\par}} 

\renewenvironment{thebibliography}[1]
        {\frenchspacing
         \ninerm\baselineskip=11pt
         \begin{list}{\arabic{enumi}.}
        {\usecounter{enumi}\setlength{\parsep}{0pt}
         \setlength{\leftmargin 12.7pt}{\rightmargin 0pt} 
         \setlength{\itemsep}{0pt} \settowidth
        {\labelwidth}{#1.}\sloppy}}{\end{list}}

\newcounter{itemlistc}
\newcounter{romanlistc}
\newcounter{alphlistc}
\newcounter{arabiclistc}

\newcommand{\fcaption}[1]{
        \refstepcounter{figure}
        \setbox\@tempboxa = \hbox{\footnotesize Fig.~\thefigure. #1}
        \ifdim \wd\@tempboxa > 5in
           {\begin{center}
        \parbox{5in}{\footnotesize\smalllineskip Fig.~\thefigure. #1}
            \end{center}}
        \else
             {\begin{center}
             {\footnotesize Fig.~\thefigure. #1}
              \end{center}}
        \fi}

\newcommand{\tcaption}[1]{
        \refstepcounter{table}
        \setbox\@tempboxa = \hbox{\footnotesize Table~\thetable. #1}
        \ifdim \wd\@tempboxa > 5in
           {\begin{center}
        \parbox{5in}{\footnotesize\smalllineskip Table~\thetable. #1}
            \end{center}}
        \else
             {\begin{center}
             {\footnotesize Table~\thetable. #1}
              \end{center}}
        \fi}

\def\@citex[#1]#2{\if@filesw\immediate\write\@auxout
        {\string\citation{#2}}\fi
\def\@citea{}\@cite{\@for\@citeb:=#2\do
        {\@citea\def\@citea{,}\@ifundefined
        {b@\@citeb}{{\bf ?}\@warning
        {Citation `\@citeb' on page \thepage \space undefined}}
        {\csname b@\@citeb\endcsname}}}{#1}}

\newif\if@cghi
\def\cite{\@cghitrue\@ifnextchar [{\@tempswatrue
        \@citex}{\@tempswafalse\@citex[]}}
\def\citelow{\@cghifalse\@ifnextchar [{\@tempswatrue
        \@citex}{\@tempswafalse\@citex[]}}
\def\@cite#1#2{{$\null^{#1}$\if@tempswa\typeout
        {IJCGA warning: optional citation argument 
        ignored: `#2'} \fi}}

\def\pmb#1{\setbox0=\hbox{#1}
        \kern-.025em\copy0\kern-\wd0
        \kern.05em\copy0\kern-\wd0
        \kern-.025em\raise.0433em\box0}

\def\fnt#1#2{\footnotetext{\kern-.3em
        {$^{\mbox{\scriptsize #1}}$}{#2}}}

\def\fpage#1{\begingroup
\voffset=.3in
\thispagestyle{empty}\begin{table}[b]\centerline{\footnotesize #1}
        \end{table}\endgroup}

\def\runninghead#1#2{\pagestyle{myheadings}
\markboth{{\protect\footnotesize\it{\quad #1}}\hfill}
{\hfill{\protect\footnotesize\it{#2\quad}}}}
\font\tenrm=cmr10
\font\tenit=cmti10 
\font\tenbf=cmbx10
\font\bfit=cmbxti10 at 10pt
\font\ninerm=cmr9

\font\eightrm=cmr8

\def\qed{\hbox{${\vcenter{\vbox{                        
   \hrule height 0.4pt\hbox{\vrule width 0.4pt height 6pt
   \kern5pt\vrule width 0.4pt}\hrule height 0.4pt}}}$}}

\renewcommand{\thefootnote}{\fnsymbol{footnote}}        

\newcommand{\abbrev}{\sf\small}
\newcommand{\api}{{\alpha_s\over \pi}}
\newcommand{\apib}{{\alpha_s^{\rm B}\over \pi}}
\newcommand{\ep}{\epsilon}
\newcommand{\eqn}[1]{Eq.\,(\ref{#1})}

\newcommand{\order}[1]{{\cal O}(#1)}
\newcommand{\bld}[1]{\boldmath{$#1$}}
\newcommand{\bsym}{\boldsymbol}

\begin{document}


\runninghead{Higgs production in gluon fusion to
  $\order{\alpha_s^4}$}{
  Higgs production in gluon fusion to
  $\order{\alpha_s^4}$
  }

\normalsize\textlineskip

\vspace*{0.88truein}

\vspace{-5em}
\begin{flushright}
  \sf BNL-HET-00/44, hep-ph/0012176 --- December 2000
\end{flushright}
\vspace{4em}

\fpage{1}
\centerline{\bf HIGGS PRODUCTION IN GLUON FUSION TO
  \bld{\order{\alpha_s^4}}\footnote{
    Talk given by RH at the {\it 
      Meeting of the Division of Particles and Fields of the 
      American Physical Society (DPF\,2000)}, 
    Columbus, Ohio, August 9--12, 2000.
    }
  }
\setcounter{footnote}{2}
\vspace*{0.37truein}
\centerline{\footnotesize ROBERT HARLANDER\footnote{email: {\tt
      rharlan@bnl.gov}}\quad and
    WILLIAM KILGORE\footnote{email: {\tt kilgore@bnl.gov}}}
\vspace*{0.015truein}
\centerline{\footnotesize\it Physics Department, Brookhaven National
  Laboratory} \baselineskip=10pt \centerline{\footnotesize\it Upton, NY
  11973, USA}

\vspace*{0.21truein}
\abstracts{The calculation of the {\footnotesize NNLO QCD} corrections
  to the partonic process $gg\to H$ is outlined.  For the coupling of
  the Higgs boson to the gluons we use an effective Lagrangian in the
  limit of a heavy top quark. The focus is on the evaluation of the
  virtual two-loop corrections. It is shown that the leading pole terms
  are in agreement with the general formula by Catani.}{}{}

\setcounter{footnote}{0}
\newpage
\thispagestyle{empty}
\mbox{}
\newpage



\runninghead{Higgs production in gluon fusion to
  $\order{\alpha_s^4}$}{
  Higgs production in gluon fusion to
  $\order{\alpha_s^4}$}

\normalsize\textlineskip
\thispagestyle{empty}
\setcounter{page}{1}

\copyrightheading{}                     

\vspace*{0.88truein}

\fpage{1}
\centerline{\bf HIGGS PRODUCTION IN GLUON FUSION TO
  \bld{\order{\alpha_s^4}}}
\vspace*{0.37truein}
\centerline{\footnotesize ROBERT HARLANDER\footnote{email: {\tt
      rharlan@bnl.gov}}\quad and
    WILLIAM KILGORE\footnote{email: {\tt kilgore@bnl.gov}}}
\vspace*{0.015truein}
\centerline{\footnotesize\it Physics Department, Brookhaven National
  Laboratory} \baselineskip=10pt \centerline{\footnotesize\it Upton, NY
  11973, USA}

\vspace*{0.21truein}
\abstracts{The calculation of the {\footnotesize NNLO QCD} corrections
  to the partonic process $gg\to H$ is outlined.  For the coupling of
  the Higgs boson to the gluons we use an effective Lagrangian in the
  limit of a heavy top quark. The focus is on the evaluation of the
  virtual two-loop corrections. It is shown that the leading pole terms
  are in agreement with the general formula by Catani.}{}{}



\textheight=7.8truein
\setcounter{footnote}{0}
\renewcommand{\thefootnote}{\alph{footnote}}


\vspace*{1pt}\textlineskip      
\section{Introduction}    
\vspace*{-0.5pt}
\noindent
Gluon fusion will be the dominant production mechanism of a Standard
Model Higgs boson at the {\abbrev LHC}.  For a Higgs mass between 100
and 200\,GeV, gluon fusion exceeds all other production channels by a
factor ranging from five to eight\cite{spirahabil}. {\abbrev QCD}
radiative corrections to this process are found to be more than 50\% at
{\abbrev NLO}\cite{dawson}.  The evaluation of the {\abbrev NNLO}
corrections ($\order{\alpha_s^4}$) is therefore highly desirable.

The first step towards the full result was taken
recently\cite{harlander}, when the virtual {\abbrev NNLO} corrections
were evaluated. Although infra-red divergent, the result could be used
to deduce an expectation of the size of the full {\abbrev NNLO}
corrections. This estimate turns out to be roughly 10--20\%,
indicating that the 
perturbative expansion is valid.

There are several more steps to be taken in order to arrive at a
prediction of the inclusive cross section at {\abbrev NNLO}. First, single and
double real radiation have to be added to the virtual corrections and
all terms must be renormalized. After
factorization of the soft singularities into the splitting functions,
this leads to an {\abbrev IR} and {\abbrev UV} finite result. Finally,
the latter has to be folded with the corresponding parton distribution
functions at {\abbrev NNLO} whose evaluation is still awaited.

In the following we will focus mainly on the virtual corrections, in
particular on the comparison of the leading poles in $\ep=(4-D)/2$ with a
general result by Catani\cite{catani} ($D$ is the space-time dimension
which is used for the regularization of the integrals).  For more
details on the actual calculation and the results we refer to
Ref.\cite{harlander}.


\section{Virtual two-loop corrections}
\noindent
It has been shown\cite{dawson} that the limit of a heavy top quark is
well justified in the process $gg\to H$. Technically this means that the
top quark loop that mediates the coupling of the gluons to the Higgs
boson can be replaced by an effective vertex. In this way, the {\abbrev
  NNLO} contribution is represented by two-loop vertex diagrams with two
massless on-shell legs. Such diagrams can be evaluated by mapping them
onto three-loop massless propagator diagrams\cite{baismi} and reducing
them through the well-known integration-by-parts algorithm\cite{chetka}.

We write the virtual contribution to the cross section for the process
$gg\to H$ as\footnote{Note that Eq.\,(9) of Ref.\cite{harlander}
  erroneously has an additional factor of $M_{\rm H}^2$.}
\begin{equation}
\begin{split}
\sigma_{\rm virt} &= {4\pi\over v^2}\left({C_1(\alpha_s)\over
    1-\beta(\alpha_s)/\ep}\right)^2{\delta(1-z)\over 256(1-\ep)}\left|1
    + {\apib} a^{(1)} 
    + \left({\apib}\right)^2 a^{(2)} + 
    \order{\alpha_s^3}
  \right|^2 =\\
  &= {4\pi\over v^2}\delta(1-z)\,{C_1^2(\alpha_s)\over 256(1-\ep)}
  \left|1
    + {\api} a_{\rm ren}^{(1)} 
    + \left({\api}\right)^2 a_{\rm ren}^{(2)} + 
    \order{\alpha_s^3}
  \right|^2\,,\\
  \beta(\alpha_s) &= -\api\beta_0 - \left(\api\right)^2\beta_1 +
    \order{\alpha_s^3}
    \,,\quad
    \beta_0 = {33 - 2n_l\over 12}\,,\quad
    \beta_1 = {153 - 19 n_l\over 24}\,.
\end{split}
\end{equation}
$C_1(\alpha_s)$ is the coefficient function for the $ggH$ vertex and
contains the residual logarithmic top mass dependence\cite{c1} ($l_t =
\ln(\mu^2/M_t^2)$, with $M_t$ the on-shell top quark mass, $n_l = 5$
denotes the number of light quark flavors, and $\alpha_s =
\alpha_s^{(5)}(\mu^2)$ is the running coupling for five active flavors)
\begin{equation}
\begin{split}
\nonumber
C_1(\alpha_s) = -{\alpha_s\over 3\pi}\left\{1 + {11\over 4}\api +
  \left(\api\right)^2\left[{2777\over 288} + {19\over 16}l_t
 + n_l\left(-{67\over 96} + {1\over 3}l_t\right)\right]\right\} + 
\order{\alpha_s^4}
\,.
\end{split}
\end{equation}
The coefficient $a^{(1)}_{\rm ren}$ up to $\order{\ep^0}$
reads\cite{dawson} ($\zeta_2 = \pi^2/6$)
\begin{equation}
\begin{split}
a^{(1)}_{\rm ren} = e^{i\pi\ep}\left({\mu^2\over M_{\rm
      H}^2}\right)^\ep\left[ -{3\over 2\ep^2} 
+ {3\over 4}\zeta_2
    \right]
+ {1\over \ep}\left(-{11\over 4} + {1\over 6}n_l\right) \,.
\label{eq::a1}
\end{split}
\end{equation}

According to Catani's general result\cite{catani}, the {\abbrev
  UV}-renormalized {\abbrev NNLO} contribution can be cast into the
following form:
\begin{equation}
\begin{split}
  a^{(2)}_{\rm ren} = {1\over 2}\bsym{I}^{(1)}(\ep)\,a^{(1)}_{\rm ren} +
  {1\over 4}\bsym{I}^{(2)}(\ep) + a^{(2)}_{\rm fin}\,,
\label{eq::a2catani}
\end{split}
\end{equation}
where $a^{(2)}_{\rm fin}$ is finite as $\ep \to 0$. In our case of two
incoming gluons, the operator $\bsym{I}^{(1)}(\ep)$ reads\footnote{ Note
  that due to a misprint ($\lambda_{ij} = +1$ instead of $\lambda_{ij} =
  -1$) the original formula\cite{catani} results in a different sign of
  the unitary phase.  A clarifying conversation on this issue with
  S.~Catani and S.~Dittmaier is acknowledged.}\,\,
($\gamma_{\rm E} = 0.577216\ldots$)
\begin{equation}
\begin{split}
\bsym{I}^{(1)}(\ep) = - \left({\mu^2\over M_{\rm H}^2}\right)^\ep 
  {e^{i\pi\ep} e^{\gamma_{\rm E}\ep}\over
  \Gamma(1-\ep)}\left[{3\over \ep^2} + {1\over \ep}
  \left({11\over 2} - {1\over 3} n_l\right)\right]\,.
\label{eq::i1}
\end{split}
\end{equation}
The expression for $\bsym{I}^{(2)}(\ep)$ is
\begin{equation}
\begin{split}
\bsym{I}^{(2)}(\ep) =& -{1\over 2}\bsym{I}^{(1)}(\ep)
\left(\bsym{I}^{(1)}(\ep) + {4\over \ep}\beta_0\right) + \\
& + {e^{-\ep\gamma_{\rm E}}\Gamma(1-2\ep)\over \Gamma(1-\ep)}
\left({2\over \ep}\beta_0 + {67\over 6} - 3\zeta_2 - {5\over 9}
  n_l\right) \bsym{I}^{(1)}(2\ep) + \bsym{H}^{(2)}(\ep)\,.
\label{eq::i2}
\end{split}
\end{equation}
$\bsym{H}^{(2)}(\ep)$ contains only single poles in $\ep$.  Thus,
Eqs.\,(\ref{eq::a1})--(\ref{eq::i2}) determine the poles of order
$1/\ep^4$, $1/\ep^3$, and $1/\ep^2$ of the {\abbrev NNLO} amplitude
$a^{(2)}_{\rm ren}$.  They can now be compared with the explicit
evaluation of the Feynman diagrams at {\abbrev NNLO}\cite{harlander},
and full agreement is found\footnote{ It is important to note that the
  prescriptions for {\scriptsize UV}-renormalization of
  Ref.\cite{catani} and Ref.\cite{harlander} differ in the treatment of
  terms $\propto \gamma_{\rm E}$ and $\propto \ln 4\pi$. For
  convenience, we apply the former prescription here.  For infra-red
  finite quantities, the two prescriptions are equivalent up to terms of
  order $\ep$.}. 
This provides a non-trivial check on the results of
Ref.\cite{harlander}. From the latter, one may now extract the
undetermined pieces $\bsym{H}^{(2)}(\ep)$ of \eqn{eq::i2} and
$a^{(2)}_{\rm fin}$ of \eqn{eq::a2catani}. 
Their values depend on whether or not one keeps the higher order terms
in $\ep$ when expanding the exponentials and the $\Gamma$-functions in
Eqs.\,(\ref{eq::a1})--(\ref{eq::i2}). Thus, in order to prevent
confusion, we refrain from giving explicit expressions for
$\bsym{H}^{(2)}(\ep)$ and $a^{(2)}_{\rm fin}$. Instead, we
advise the interested reader to use the results of
Ref.\cite{harlander}.

\section{Real radiation}
\noindent
Let us give a brief description of the contributions coming from the
real emission of quarks and gluons.  The divergences of the virtual
two-loop corrections will be canceled by the soft contributions of the
following processes: $gg\to Hg$ to one-loop order, and $gg\to Hgg$, $gg
\to Hq\bar q$ at tree level (assuming the effective $ggH$ vertex).
After adding these contributions to the virtual corrections, there are
still infra-red singularities left. These have to be absorbed by the
mass factorization procedure, using the Altarelli-Parisi splitting
functions at {\abbrev NNLO}.

Explicit results for the full partonic calculation will be given
elsewhere\cite{ggH}. For the contribution $\propto \delta(1-s/M_{\rm
  H}^2)$ ($s$ is the cms energy of the $gg$-system) we find a {\abbrev
  NNLO} correction of order 10\%, in agreement with the numerical
estimate of Ref.\cite{harlander}.


\nonumsection{Acknowledgements}
\noindent
The work of RH is supported by {\it Deutsche Forschungsgemeinschaft}.


\def\app#1#2#3{{\it Act.~Phys.~Pol.~}{\bf B #1} (#2) #3}
\def\apa#1#2#3{{\it Act.~Phys.~Austr.~}{\bf#1} (#2) #3}
\def\annphys#1#2#3{{\it Ann.~Phys.~}{\bf #1} (#2) #3}
\def\cmp#1#2#3{{\it Comm.~Math.~Phys.~}{\bf #1} (#2) #3}
\def\cpc#1#2#3{{\it Comp.~Phys.~Commun.~}{\bf #1} (#2) #3}
\def\epjc#1#2#3{{\it Eur.\ Phys.\ J.\ }{\bf C #1} (#2) #3}
\def\fortp#1#2#3{{\it Fortschr.~Phys.~}{\bf#1} (#2) #3}
\def\ijmpc#1#2#3{{\it Int.~J.~Mod.~Phys.~}{\bf C #1} (#2) #3}
\def\ijmpa#1#2#3{{\it Int.~J.~Mod.~Phys.~}{\bf A #1} (#2) #3}
\def\jcp#1#2#3{{\it J.~Comp.~Phys.~}{\bf #1} (#2) #3}
\def\jetp#1#2#3{{\it JETP~Lett.~}{\bf #1} (#2) #3}
\def\mpl#1#2#3{{\it Mod.~Phys.~Lett.~}{\bf A #1} (#2) #3}
\def\nima#1#2#3{{\it Nucl.~Inst.~Meth.~}{\bf A #1} (#2) #3}
\def\npb#1#2#3{{\it Nucl.~Phys.~}{\bf B #1} (#2) #3}
\def\nca#1#2#3{{\it Nuovo~Cim.~}{\bf #1A} (#2) #3}
\def\plb#1#2#3{{\it Phys.~Lett.~}{\bf B #1} (#2) #3}
\def\prc#1#2#3{{\it Phys.~Reports }{\bf #1} (#2) #3}
\def\prd#1#2#3{{\it Phys.~Rev.~}{\bf D #1} (#2) #3}
\def\pR#1#2#3{{\it Phys.~Rev.~}{\bf #1} (#2) #3}
\def\prl#1#2#3{{\it Phys.~Rev.~Lett.~}{\bf #1} (#2) #3}
\def\pr#1#2#3{{\it Phys.~Reports }{\bf #1} (#2) #3}
\def\ptp#1#2#3{{\it Prog.~Theor.~Phys.~}{\bf #1} (#2) #3}
\def\ppnp#1#2#3{{\it Prog.~Part.~Nucl.~Phys.~}{\bf #1} (#2) #3}
\def\sovnp#1#2#3{{\it Sov.~J.~Nucl.~Phys.~}{\bf #1} (#2) #3}
\def\tmf#1#2#3{{\it Teor.~Mat.~Fiz.~}{\bf #1} (#2) #3}
\def\tmp#1#2#3{{\it Theor.~Math.~Phys.~}{\bf #1} (#2) #3}
\def\yadfiz#1#2#3{{\it Yad.~Fiz.~}{\bf #1} (#2) #3}
\def\zpc#1#2#3{{\it Z.~Phys.~}{\bf C #1} (#2) #3}
\def\ibid#1#2#3{{ibid.~}{\bf #1} (#2) #3}

\end{document}